\documentstyle[12pt,epsfig]{article}  
\begin{document} 
\baselineskip=20pt

\def\la{\mathrel{\mathpalette\fun <}}
\def\ga{\mathrel{\mathpalette\fun >}}
\def\fun#1#2{\lower3.6pt\vbox{\baselineskip0pt\lineskip.9pt
\ialign{$\mathsurround=0pt#1\hfil##\hfil$\crcr#2\crcr\sim\crcr}}} 

\begin{titlepage} 
\begin{center}
{\Large \bf Partonic energy loss in ultrarelativistic heavy ion collisions: jet  
suppression versus jet fragmentation softening} \\

\vspace{4mm}

I.P.~Lokhtin and A.M.~Snigirev  \\
M.V.Lomonosov Moscow State University, D.V.Skobeltsyn Institute of Nuclear Physics \\
119992, Vorobievy Gory, Moscow, Russia \\ E:mail:~~igor@lav01.sinp.msu.ru,
~~snigirev@lav01.sinp.msu.ru  
\end{center}  

\begin{abstract} 
We discuss the modification of a jet fragmentation function due to medium-induced 
partonic energy loss in context of leading particle observables in ultrarelativistic
nucleus-nucleus interactions. We also analyze the relation between in-medium softening 
jet fragmentation function and suppression of the jet rates due to energy loss outside 
the jet cone. The predicted anti-correlation between two effects allows to probe 
a fraction of partonic energy loss carried out of the jet cone and truly lost to the 
jet. 
\end{abstract}

\bigskip

\noindent 
$PACS$: ~~12.38.Mh, 24.85.+p, 25.75.+r \\ 
$Keywords$:~~jet energy loss, leading particles, jet fragmentation, relativistic 
nuclear collisions \\
\end{titlepage}   

\section{Introduction} 

Jets play one of the central roles as a promising tool to study properties of 
quark-gluon plasma (QGP) expected to be created in heavy ion collisions at RHIC and 
LHC. Medium-induced energy loss of energetic partons, the so-called jet quenching, has 
been proposed to be very different in cold nuclear matter and in QGP, resulting in many 
challenging observable phenomena (see for the review~\cite{baier_rev} and references 
therein). In particular, softening jet fragmentation function and, as a consequence,  
suppression of high-$p_T$ hadron spectrum in nucleus-nucleus collisions relative to 
their production in independent nucleon-nucleon interactions, are 
considered~\cite{Gyulassy:1992,Baier:2001,Gyulassy:2002,Wang:2002,Salgado:2002, 
Muller:2002,Wiedemann:2003}. 
Recent RHIC data on inclusive high-p$_T$ charge
and neutral hadron production from STAR~\cite{star}, PHENIX~\cite{phenix},
PHOBOS~\cite{phobos} and BRAHMS~\cite{brahms} experiments show such kind of
suppression and are in agreement with the jet quenching hypothesis. However, since
at the moment direct event-by-event reconstruction of jets and their characteristics 
is not available in RHIC experiments, the assumption that integrated yield of all 
high-$p_T$ particles originates only from jet fragmentation is not fully 
clear (see, for example,~\cite{Shuryak:2002}). At the LHC, a new regime is reached 
where hard and semi-hard QCD multi-particle production can certainly dominate over 
underlying soft events~\cite{lhc}. CMS experiment at LHC~\cite{cms94} will be able to 
provide adequate jet reconstruction using calorimetric
measurements~\cite{note00-060,lhc}. Thus identification of leading particle in a jet 
(i.e. particle carrying the maximal fraction of jet transverse momentum) allows the
measurement of jet fragmentation function (JFF) to be done. Comparison of JFF in $AA$
and $pp$ collisions (or in central and peripheral $AA$ interactions) may give 
information about in-medium modification of JFF. 

The crucial related question here is: how much energy loss falls {\it outside}
the typical jet cone and is truly lost to the jet? There are some discussions on 
this subject in the
literature~\cite{lokhtin98,baier,Zakharov:1999,urs,vitev}. In fact, since coherent 
Landau-Pomeranchuk-Migdal radiation induces a strong dependence of the radiative energy 
loss of a jet on the angular cone size, it will soften particle energy 
distributions inside the jet, increase the multiplicity of secondary particles, and to 
a lesser degree, affect the total jet energy. On the other hand, collisional energy 
loss turns out to be practically independent of jet cone size and causes the 
loss of total jet energy, because the bulk of ``thermal'' particles knocked out of the 
dense matter by elastic scatterings fly away in almost transverse direction relative to 
the jet axis~\cite{lokhtin98}. Moreover, the total energy loss of a jet will be 
sensitive to the experimental capabilities for low-p$_T$ particles, products of soft 
gluon fragmentation. For example, in the CMS case, most of these low-$p_T$ hadrons may 
be cleared out of the central calorimeters by the strong magnetic 
field~\cite{cms94,note00-060}.

In this Letter we analyze relation between in-medium softening JFF and suppression of 
jet spectrum due to energy loss outside the jet cone. 

\section{Medium-modified jet fragmentation}

Let us recall that in the leading order of perturbative QCD the jet production cross 
section with transverse momentum $p_T$ and rapidity $y$ in a nucleon-nucleon collision 
is given by 
\begin{equation}
\frac{d\sigma^{\rm jet(k)}}{dp_T^2dy} = \sum_{ij} \int dx_i  dx_j  
f_a^i(x_i, Q^2) f_b^j(x_j, Q^2) \frac{d\widehat{\sigma }}
{d\widehat{t}}(ij\rightarrow kl) \delta (\xi -1), 
\end{equation}  
where $x_i=p_i/p_a$, $x_j=p_j/p_b$ are the initial momentum fractions of nucleons   
$a$, $b$ carried by
the interacting partons of types $i$, $j$; $f_a^i(x_i, Q^2)$, $f_b^j(x_j, Q^2)$ are
the parton distribution functions for the colliding nucleons $a$, $b$;   
$\widehat s$, $\widehat t$ and $\widehat u$ are the Mandelstam variables of hard
parton subprocess; $d\widehat{\sigma }/d\widehat t (ij\rightarrow kl)$ is the Born cross section
for the hard $ij\rightarrow kl$ scattering subprocess;
$Q^2=p_T^2=(\widehat{t}\widehat{u})/\widehat{s}$;
$\widehat{t}/\widehat{u}=(x_i/x_j)\exp{(-2y)}$; 
$\xi=p_T/\sqrt{s}(\exp{(y)}/x_i+\exp{(-y)}/x_j)$ is the momentum fraction of parton $k$ 
carried by the jet; $\sqrt{s}=\sqrt{\widehat{s}/(x_i x_j)}$ is the center of mass 
energy of the colliding nucleons.  
To calculate the inclusive cross
section for ``jet-induced'' hadron production with transverse momentum $p_T^h$ and 
rapidity $y^h$ in a nucleon-nucleon collision, one convolutes the jet production cross 
section with the fragmentation function $D_k^h (z', p_T^2)$ for the parton of type $k$ 
into hadron $h$:    
\begin{equation}
\frac{d\sigma^{\rm h(k)}}{d(p_T^h)^2dy^hdz'} = \frac{d\sigma^{\rm jet(k)}}{dp_T^2dy}
\frac{1}{z'^2} D_k^h (z', p_T^2) ,  
\end{equation}
where $y^h=y$, and $z'=p^h/p^k=p_T^h/p_T$ is the momentum fraction of a parton $k$  
carried by the final observable hadron. 

The parton distribution functions $f_a^i(x_i, Q^2)$ are measured in deep inelastic
scattering experiments such as those at HERA~\cite{hera}, while fragmentation
functions $D_k^h (z', Q^2)$ are extracted from $e^+e^-$ annihilation from PETRA, PEP
and LEP~\cite{lep} and from hadronic collisions from UA1~\cite{ua1} (gluon
fragmentation function).  

In nuclear interactions medium-induced energy loss of fast partons can modify the
cross sections for high-$p_T$ hadrons and jets together with other potentially
important nuclear effects like, for instance, parton shadowing playing significant
role at small values of the momentum fraction carried by the interacting partons.
However for sufficiently hard jets and hadrons under consideration ($x_{i,j}\ga 0.2$) 
this effect is negligible as well as next-to-leading order (NLO) corrections ($K$
factor $\sim 1$)~\cite{lokhtin00}. As a result the rate of $k$-type jets of finite 
angular cone size $\theta_0$ in mid-rapidity with transverse momentum $p_T^{\rm jet}$ 
in $AA$ collisions at given impact parameter $b$ can be estimated as 
\begin{equation}
\label{vertex}
\frac{dN_{AA}^{\rm jet(k)}}{d(p_T^{\rm jet})^2dy} (\theta_0,b) = \int\limits_0^{2\pi} d \psi 
\int\limits_0^{r_{max}}r dr T_A(r_1) T_A(r_2)   
\frac{d\sigma^{\rm jet(k)}(p_T^{\rm jet}
+\Delta p_T^{\rm jet}(r, \psi, \theta_0))}{dp_T^2dy} , 
\end{equation} 
where $r_{1,2} (b,r,\psi)$ are the distances between the nucleus centers and the jet
production vertex $V(r\cos{\psi}, r\sin{\psi})$; $r_{max} (b, \psi) \le R_A$ is the 
maximum possible transverse distance $r$ from the nuclear collision axis to the $V$; 
$R_A$ is the radius of the nucleus $A$; $T_A(r_{1,2})$ is the nuclear 
thickness function (see Ref.~\cite{lokhtin00} for detailed nuclear geometry 
explanations). The effective shift $\Delta p_T^{\rm jet} (r, \psi, \theta_0)$ of  
jet momentum spectrum depends on the jet cone size $\theta_0$.  

The partons will not hadronize inside QGP. In a hadronic medium, we assume that the
fragmentation functions can be approximated by their forms in vacuum (see, however,
Refs.~\cite{Zakharov:2002,Osborne:2002} where the corrections to the parton fragmentation 
function in a thermal medium were discussed).  We take into consideration the
fragmentation of leading partons only and omit the fragmentation of emitted
gluons, because we are interested here in the leading particle in a jet. Then the
rate of high-$p_T$ jet-induced hadrons can be estimated as
\begin{equation}
\frac{dN_{AA}^{\rm h(k)}}{d(p_T^h)^2dydz'} (b) = \int\limits_0^{2\pi} d \psi 
\int\limits_0^{r_{max}}r dr T_A(r_1) T_A(r_2) 
\frac{d\sigma^{\rm jet(k)}(p_T+\Delta p_T(r, \psi))}{dp_T^2dy} \frac{1}{z'^2}   
D^h_k(z', p_T^2)~,   
\end{equation} 
where the shift $\Delta p_T$ of hadron momentum distribution generally is not equal to 
the mean in-medium partonic energy loss due to the steep fall-off of the 
$p_T$-spectrum~\cite{Baier:2001}.

The integral jet suppression factor $Q$ can be introduced by the natural way as the
ratio of jet rate with energy loss to jet rate without one, 
\begin{equation}
\label{qq}
Q^{\rm jet}(p_{T~{\rm min}}^{\rm jet})=\int\limits_{p_{T~{\rm min}}^{\rm jet}}
 d(p_T^{\rm jet})^2dy \frac{dN_{AA}^{\rm jet(k)}}{d(p_T^{\rm jet})^2dy} 
\Bigg/ \int\limits_{p_{T~{\rm min}}^{\rm jet}} d(p_T^{\rm jet})^2dy 
\frac{dN_{AA}^{\rm jet(k)}}{d(p_T^{\rm jet})^2dy} 
(\Delta p_T^{\rm jet}=0) . 
\end{equation} 
JFF is defined here as 
\begin{equation}
\label{dz}
D(z)=\int \limits_{z\cdot p_{T~{\rm min}}^{\rm jet}} d(p^h_T)^2 dy dz'
\frac{dN_{AA}^{\rm h(k)}}{d(p^h_T)^2dydz'}\delta(z-p_T^h/p_T^{\rm jet})
 \Bigg/ \int \limits_{p_{T~{\rm
min}}^{\rm jet}} d(p_T^{\rm jet})^2 dy 
\frac{dN_{AA}^{\rm jet(k)}}{d(p_T^{\rm jet})^2dy}  
\end{equation}
and coincides approximately with $D_k^h (z, (p_{T~{\rm min}}^{\rm jet})^2)$ for the 
case without energy loss and with the type of jet specified.
Note that $z \equiv p_T^h/p_T^{\rm jet}$($=
 z'p_T/p_T^{\rm jet}$) is 
experimentally observable quantity depending on jet cone size $\theta_0$. 
Further we are interested also in the ratio 
\begin{equation}
\label{dd}
\frac{D(z>z_0)}{D(z>z_0, \Delta p_T=0)} \equiv \frac{\int \limits_{z_0}^1 dz D(z)}{
\int \limits_{z_0}^1 dz D(z, \Delta p_T=0)}=\frac{Q^h(p_{T~{\rm min}}^{\rm
jet}, z_0)}{Q^{\rm jet}(p_{T~{\rm min}}^{\rm jet})}~,
\end{equation}
where the integral hadron suppression factor 
\begin{eqnarray}
\label{qqq}
Q^h(p_{T~{\rm min}}^{\rm jet}, z_0) & = & \int \limits_{z_0}^1 dz\int \limits_{z\cdot 
p_{T~{\rm min}}^{\rm jet}} d(p^h_T)^2 dydz' 
\frac{dN_{AA}^{\rm h(k)}}{d(p^h_T)^2dydz'} \delta(z-p_T^h/p_T^{\rm jet})
\Bigg/ \nonumber \\ & & 
\int \limits_{z_0}^1 dz\int \limits_{z\cdot p_{T~{\rm min}}^{\rm jet}}
 d(p^h_T)^2 dydz' 
\frac{dN_{AA}^{\rm h(k)}}{d(p^h_T)^2dydz'} (\Delta p_T = 0)\delta(z-p_T^h/p_T^{\rm jet})
\end{eqnarray}
is distinguished from the differential hadron quenching factor usually defined 
as~\cite{Baier:2001,Salgado:2002,Muller:2002,Wiedemann:2003} 
\begin{equation}
\label{qqqq}
\bar{Q^h}(p_T^h)=\int dy dz' \frac{dN_{AA}^{\rm h(k)}}{d(p^h_T)^2dydz'} \Bigg/ \int dy 
dz' \frac{dN_{AA}^{\rm h(k)}}{d(p^h_T)^2dydz'} (\Delta p_T = 0) . 
\end{equation}

\section{Model} 

In order to generate the initial jet distributions in nucleon-nucleon sub-collisions at 
$\sqrt{s}=5.5$ TeV, we have used PYTHIA$\_5.7$~\cite{pythia}. After that we perform
event-by-event Monte-Carlo simulation of rescattering and energy loss of partons in QGP 
(for model details one can refer to our previous papers~\cite{lokhtin98,lokhtin00}). 
The approach relies on an accumulative energy losses, when gluon radiation is 
associated with each scattering in expanding medium together including the interference 
effect by the modified radiation spectrum as a function of decreasing temperature 
$dE/dl(T)$. The basic kinetic integral equation for the energy loss $\Delta E$ as a 
function of initial energy $E$ and path length $L$ has the form 
\begin{eqnarray} 
\label{elos_kin}
\Delta E (L,E) = \int\limits_0^Ldl\frac{dP(l)}{dl}
\lambda(l)\frac{dE(l,E)}{dl} \, , ~~~~ 
\frac{dP(l)}{dl} = \frac{1}{\lambda(l)}\exp{\left( -l/\lambda(l)\right) }
\, ,  
\end{eqnarray} 
where $l$ is the current transverse coordinate of a parton, $dP/dl$ is the scattering
probability density, $dE/dl$ is the energy loss per unit length, $\lambda = 1/(\sigma 
\rho)$ is in-medium mean free path, $\rho \propto T^3$ is medium density at temperature 
$T$, $\sigma$ is the integral cross section of parton interaction in the medium.
Such numerical simulation of free path of a hard jet in QGP allows us to obtain any 
kinematical characteristic distributions of jets in final state. Besides the different 
scenarios of medium evolution can be considered. 

For the calculations we have used collisional part of loss~\cite{lokhtin00},  
\begin{equation} 
\label{col} 
\frac{dE}{dl}^{col} = \frac{1}{4T \lambda \sigma} 
\int\limits_{\displaystyle\mu^2_D}^
{\displaystyle 3T E / 2}dt\frac{d\sigma }{dt}t ~,
\end{equation} 
and the dominant contribution to the differential cross section 
\begin{equation} 
\frac{d\sigma }{dt} \cong C \frac{2\pi\alpha_s^2(t)}{t^2} ~,~~~~
\alpha_s = \frac{12\pi}{(33-2N_f)\ln{(t/\Lambda_{QCD}^2)}} \>
\end{equation} 
for scattering of a parton with energy $E$ off the ``thermal'' partons with energy (or 
effective mass) $m_0 \sim 3T \ll E$. Here $C = 9/4, 1, 4/9$ for $gg$, $gq$ and 
$qq$ scatterings respectively, $\alpha_s$ is the QCD running coupling constant for 
$N_f$ active quark flavours, and 
$\Lambda_{QCD}$ is the QCD scale parameter which is of the order of the critical 
temperature,  $\Lambda_{QCD}\simeq T_c \simeq 200$ MeV. The integrated cross 
section $\sigma$ is regularized by the Debye screening mass squared $\mu_D^2 (T)$. 

The energy spectrum of coherent medium-induced gluon radiation and the corresponding
dominated part of radiative energy loss was estimated using BDMS formalism~\cite{baier}: 
\begin{eqnarray} 
\label{radiat} 
\frac{dE}{dl}^{rad} = \frac{2 \alpha_s (\mu_D^2) C_R}{\pi L}
\int\limits_{\omega_{\min}}^E  
d \omega \left[ 1 - y + \frac{y^2}{2} \right] 
\>\ln{\left| \cos{(\omega_1\tau_1)} \right|} 
\>, \\  
\omega_1 = \sqrt{i \left( 1 - y + \frac{C_R}{3}y^2 \right)   
\bar{\kappa}\ln{\frac{16}{\bar{\kappa}}}}
\quad \mbox{with}\quad 
\bar{\kappa} = \frac{\mu_D^2\lambda_g  }{\omega(1-y)} ~, 
\end{eqnarray} 
where $\tau_1 = L / (2 \lambda_g)$, $y = \omega / E$ is the fraction of the 
hard parton energy carried by the radiated gluon, and $C_R = 4/3$ is the quark colour 
factor. A similar expression for the gluon jet can be obtained by substituting 
$C_R=3$ and a proper change of the factor in the square bracket in (\ref{radiat}), see
Ref.~\cite{baier}. The integral (\ref{radiat}) is carried out over all energies from 
$\omega_{\min}=E_{LPM}=\mu_D^2\lambda_g$ ($\lambda_g$ is the gluon mean free path), the 
minimal radiated gluon energy in the coherent LPM regime, up to initial jet energy $E$. 
Note that although the radiative energy loss of an energetic parton dominates over the 
collisional loss by up to an order of magnitude, the relative contribution of
collisional loss of a jet growths with increasing jet cone size due to essentially 
different angular structure of loss for two mechanisms~\cite{lokhtin98}. 

The medium was treated as a boost-invariant longitudinally expanding quark-gluon fluid, 
and partons as being produced on a hyper-surface of equal proper times 
$\tau$~\cite{bjorken}. For certainty we used the initial conditions 
for the gluon-dominated plasma formation 
expected for central Pb$-$Pb collisions at LHC~\cite{esk}: 
$\tau_0 \simeq 0.1$ fm/c, $T_0 \simeq 1$ GeV, $\rho_g \approx 
1.95T^3$. For non-central collisions we suggest the proportionality of the initial 
energy density to the ratio of nuclear overlap function 
and effective transverse area of nuclear overlapping~\cite{lokhtin00}.

Thus we consider that in each $i$-th scattering off the comoving particle (with the
same longitudinal rapidity) a fast parton loses energy 
collisionally and radiatively, $\Delta e_i = t_i/(2m_0) + \omega _i$, where $t_i$ and
$\omega _i$ are simulated according to Eqs.~(\ref{col}) and (\ref{radiat})
respectively, and the distribution of jet production vertex -- according to  
Eq.~(\ref{vertex}). Finally we suppose that in every event the energy of an initial parton decreases by 
value $\Delta p_T (r, \psi )= \sum _i \Delta e_i$ and the jet energy loss, 
$\Delta p_T^{\rm jet}$, is the product of $\varepsilon$ times energy loss of an initial 
parton, $\Delta p_T^{\rm jet} (r,\psi,\theta_0) = \varepsilon \cdot \Delta p_T (r,\psi )$. 
We examine here the fraction $\varepsilon$ of partonic energy loss carried out of the 
jet cone as a phenomenological parameter, since the treatment of angular spectrum of 
emitted gluons is rather sophisticated and 
model-dependent~\cite{lokhtin98,baier,Zakharov:1999,urs,vitev}~\footnote{Of course, 
this parameter is regulated by the jet cone size $\theta_0$ and its
dependence on $\theta_0$ was numerically investigated~\cite{baier,urs}. We can
formally recalculate our result as a function of $\theta_0$ using such numerical 
estimation or taking into consideration the angular structure of radiative and
collisional energy loss in each scattering complicating our simulation. However this
analysis demands also the jet finding procedure specification and other technical
details. We believe that our simplified treatment here is enough to demonstrate the 
anti-correlation between jet suppression and JFF softening, moreover the integral 
hadron suppression factor $Q^h(p_{T~{\rm min}}^{\rm jet},z_0)$ in Eq.~(\ref{dd}) 
depends on $\varepsilon$ implicitly (only via $z$-definition).}. We was interested in 
relatively realistic values of $\varepsilon$ in the range from $0$ to $1$ (although 
$\varepsilon$ can be even larger than $1$ at small $\theta_0$, see~\cite{baier,urs}).  

In our simulations we have considered channel with neutral pions only as leading
particles in jets. One can expect the similar results for leading
charged hadrons. From methodical point of view tracking in
heavy ion environment at LHC is rather complicated (although solvable)
task, while the reconstruction of electromagnetic clusters in
calorimeters being at the moment more understandable~\cite{lhc,note00-060}. 
At high enough transverse momentum of
$\pi_0$ ($\ga 15$ GeV at CMS case~\cite{lhc}), two photons from pion
decay fall into one crystal of electromagnetic calorimeter, and 
traditional technique for reconstructing $\pi_0$'s using two-photon
invariant mass spectrum does not work. However, such electromagnetic
cluster can be identified as a leading $\pi_0$, if it belongs a hard
jet and carries the significant part of jet transverse energy. 

\section{Numerical results and conclusions} 

Figure 1 shows JFF~(\ref{dz}) (without the jet-type specification and therefore
experimentally observable) for leading $\pi_0$'s for the cases without 
and with medium-induced energy loss obtained in the frameworks of our
model~\cite{lokhtin98,lokhtin00} in central and minimum-bias Pb$-$Pb collisions. 
The threshold for jet reconstruction 
$p_{T~\rm min}^{\rm jet} = 100$ GeV was used~\cite{note00-060}~\footnote{The
estimated statistics for $>100$ GeV jets in CMS acceptance is 
high enough, at the level $\sim 10^7$ jet pairs per $1$ month of LHC 
Pb$-$Pb run~\cite{note00-060,lhc}}. If $\varepsilon$ close to $0$ (``small-angular 
radiation'' dominates), then the factor of jet suppression $Q^{\rm jet}$~(\ref{qq}) 
is close to $1$ (there is almost no jet rate suppression), and effect on JFF softening 
is maximal. Increasing $\varepsilon$ (the contribution from 
``wide-angular radiation'' and collisional loss grows) results in 
stronger jet suppression ($Q^{\rm jet}$ value decreases), but effect on JFF
softening becomes smaller, especially for highest $z$. Figure 2 presents the
$\varepsilon$-dependences for jet suppression factor $Q^{\rm jet}$ 
(without the jet-type specification) and 
ratio (\ref{dd}) of JFF with energy loss to JFF without loss, 
$D (z>z_0) / D (z>z_0, \Delta p_T=0)$, for $z_0=0.5$ and $0.7$. Note that
in the case without jet quenching, fraction of events when
leading $\pi_0$ carries larger $50\%$ ($70\%$) of jet transverse momentum
is $6.3 \cdot 10^{-3}$ ($9.3 \cdot 10^{-4}$). One can see the 
distinctive anti-correlation between strengthening jet suppression and 
JFF softening (ratio (\ref{dd}) can be even greater than $1$ at large enough 
$\varepsilon$ and $z$ values) determined mainly by fraction 
of partonic energy loss carried outside the jet cone. The physical reason for the
effect to be opposite in the jet suppression factor and the fragmentation function is
it follows. Increasing $\varepsilon$ results in decreasing final jet transverse
momentum, $p_T^{\rm jet}=p_T^{\rm jet}(\Delta p_T^{\rm jet}=0)-\varepsilon \cdot \Delta 
p_T^{\rm jet}$ (which is the denominator in definition of $z\equiv p_T^h/p_T^{\rm jet}$ 
in JFF~(\ref{dz})) without any 
influence on the numerator of $z$ and, as a consequence, in reducing effect 
on JFF softening, while the integral jet suppression
factor~(\ref{qq}) becomes larger. The remarkable prediction here 
is that the effect on jet rate suppression becomes comparable with the effect on JFF 
softening at quite reasonable value $\varepsilon \sim 0.3$. 

In summary, we have analyzed the relation between in-medium softening 
jet fragmentation function (in leading $\pi_0$ channel) and suppression of the 
jet spectrum due to energy loss outside the jet cone. 
We believe that this kind of analysis can be performed for the  
heavy ion collisions at LHC experiments. The observation of 
significant JFF softening without substantial jet rate suppression 
would be an indication of the fact that small-angular gluon radiation is 
dominating mechanism for medium-induced partonic energy loss. Increasing
contribution of wide-angular gluon radiation and collisional loss can
result in jet rate suppression, while the effect on JFF softening becomes less in this
case. If the contribution of the ``out-of-cone'' jet energy loss is large enough, the 
jet rate suppression may be even more significant than JFF softening.  

{\it Acknowledgements.}  
Discussions with Yu.L.~Dokshitzer, L.I.~Sarycheva, I.N.~Vardanian, U.~Wiedemann, 
P.~Yepes, B.~Zakharov and G.M.~Zinovjev are gratefully acknowledged.

\begin{figure}[hbtp] 
\begin{center} 
\makebox{\epsfig{file=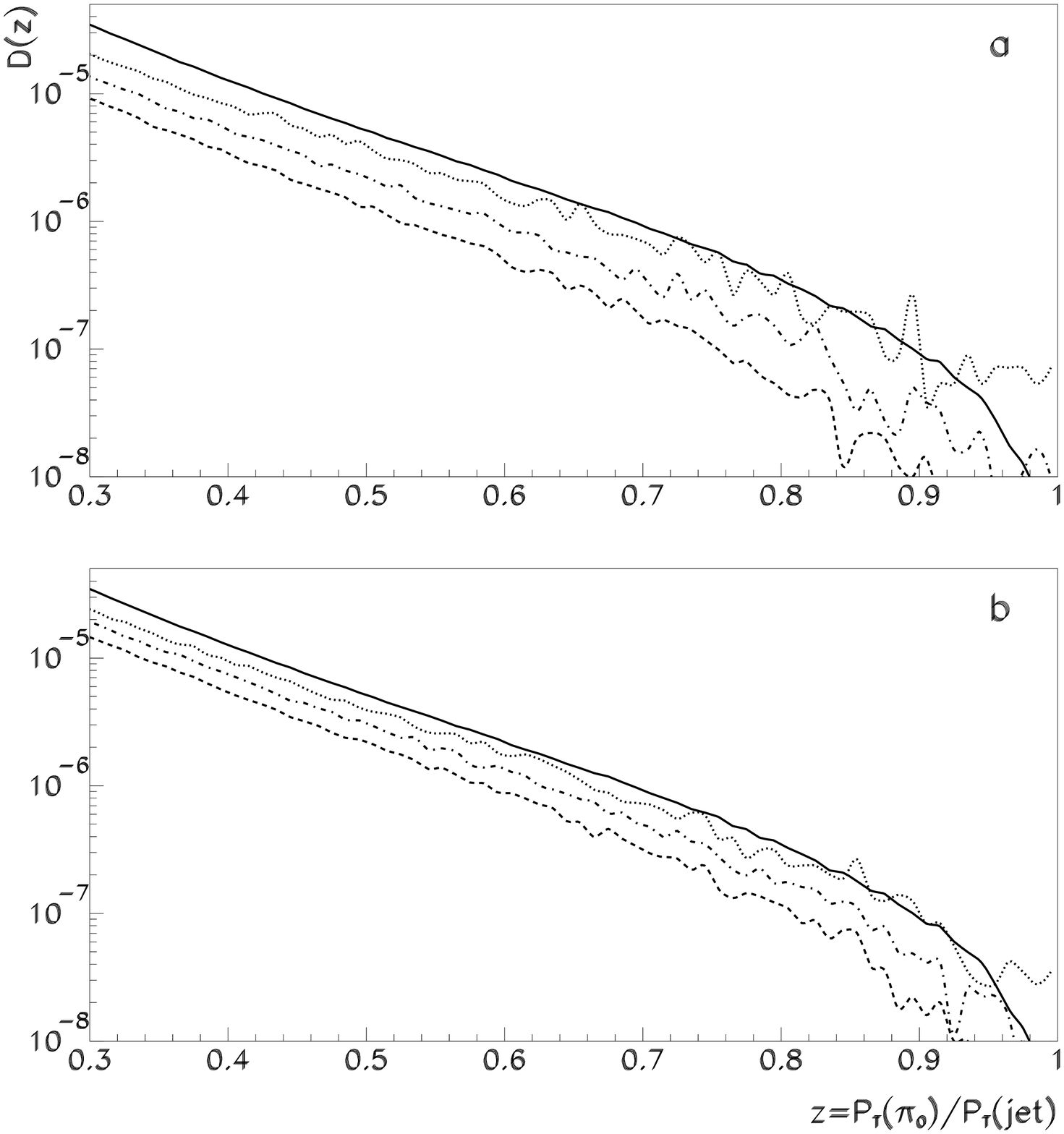, height=200mm}}   
\caption{Jet fragmentation function for leading $\pi_0$'s without (solid curves) and 
with medium-induced energy loss for $\varepsilon=0$ (dashed curves), $0.3$
(dash-dotted curves) and $0.7$ (dotted curves) in central ({\em a}) and minimum-bias 
({\em b}) Pb$-$Pb collisions. $p_T^{\rm jet}>100$ GeV and $\mid y^{jet}\mid < 3$.}    
\end{center}
\end{figure}

\begin{figure}[hbtp] 
\begin{center} 
\makebox{\epsfig{file=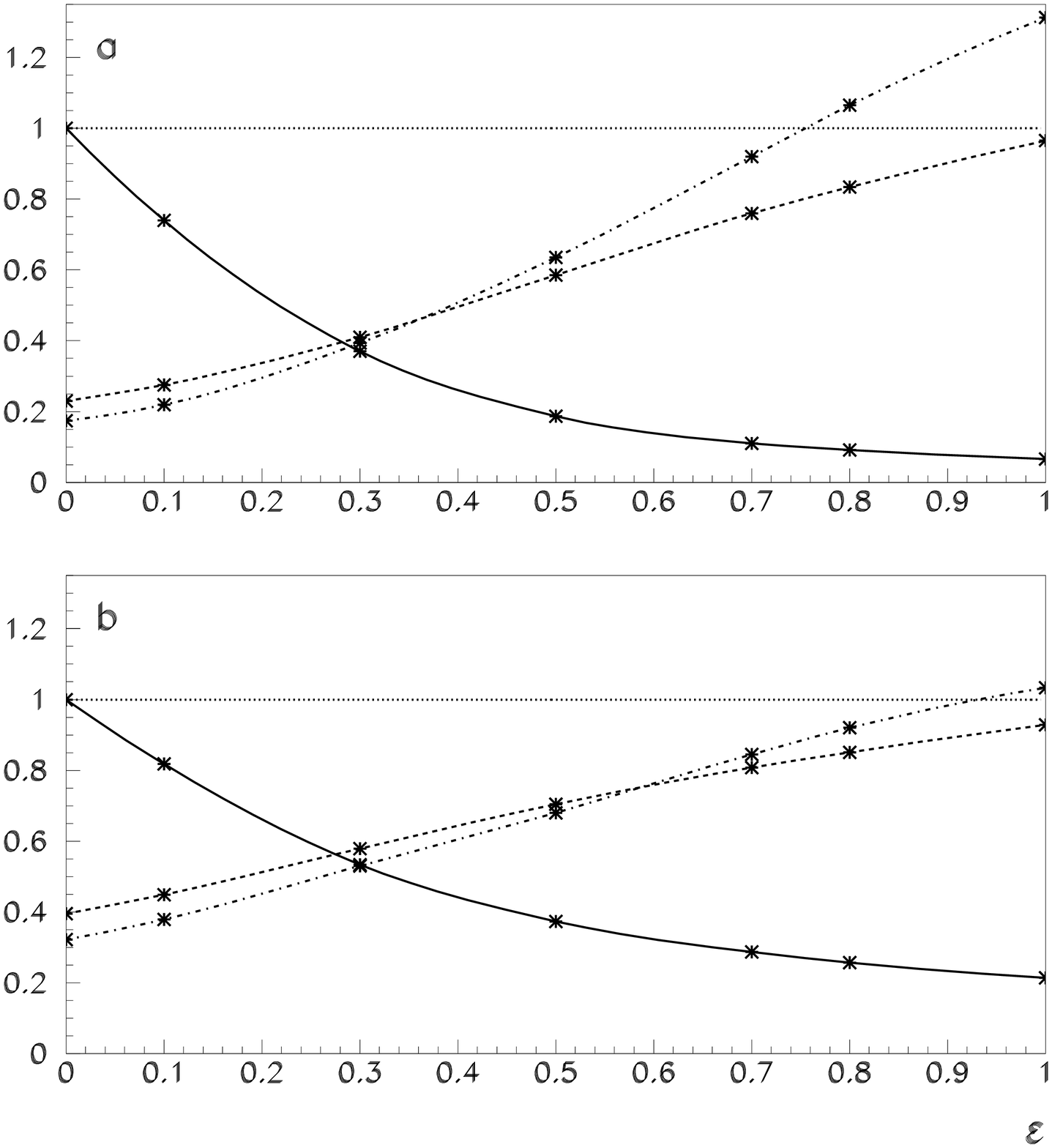, height=200mm}}   
\caption{Jet suppression factor $Q^{\rm jet}$ (solid curves) and ratio of JFF with 
energy loss to JFF without loss, $D (z>z_0) / D (z>z_0, \Delta p_T=0)$, for $z_0=0.5$ (dashed
curves) and $0.7$ (dash-dotted curves) in central ({\em a}) and minimum-bias 
({\em b}) Pb$-$Pb collisions. $p_T^{\rm jet}>100$ GeV and $\mid y^{jet}\mid < 3$.}   
\end{center}
\end{figure}

\end{document}